# The Interaction of Matter and Radiation:
# The Physics of C.V. Raman, S.N. Bose and M.N. Saha
# Part 1: Historical Background

Arnab Rai Choudhuri
Department of Physics
Indian Institute of Science
Bengaluru – 560012
e-mail: arnab@iisc.ac.in

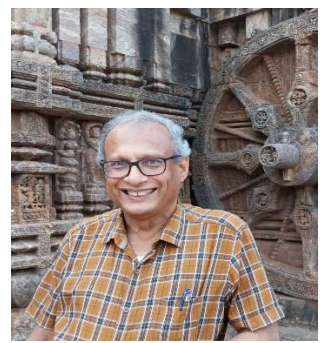

The author is a theoretical astrophysicist working at the Department of Physics, Indian Institute of Science. He was involved in developing the currently accepted theoretical model of the 11-year sunspot cycle. He also has interest in the history and philosophy of science.

**Summary.** Three extraordinary physics discoveries were made from colonial India, which did not have any previous tradition of research in modern physics: Saha ionization equation (1920), Bose statistics (1924), Raman effect (1928). All the three discoverers were founding faculty members of the new small physics department of Calcutta[1] University, which started functioning from 1916. These discoveries were all in the general topic of interaction between matter and radiation. In Part 1 of this article, we describe the social and the intellectual environment in which these discoveries were made. Part 2 will focus on the science involved in these discoveries.

**Keywords:** Culcutta University – Raman effect – Bose(-Einstein) statistics – Saha ionization equation

## Introduction

Any well-informed Indian with some interest in scientific matters would know that Chadrasekhara Ventata Raman (1888 – 1970), Satyendranath Bose (1894 – 1974) and Meghnad Saha (1893 – 1956), working under very adverse conditions in colonial India, reached extraordinary heights of scientific creativity in physics approximately a century before the present time. However, apart from professional physicists who have to learn about their works as part of university-level physics curriculum, very few non-physicists have much idea of what exactly Raman, Bose and Saha achieved and how important really their works were in the development of physics.  The aim of Part 1 of this article is to discuss the circumstances under which these works were done.  Part 2 will then give non-technical accounts of these works and their significance, assuming only a very rudimentary acquaintance with physics.

---

[1] Since this article is of historical nature, I have followed the standard convention of historical scholarship in writing the names of places in the way they were written in the early decades of the twentieth century.

An extraordinary scientific revolution based on relativity and quantum theory – perhaps the most important scientific revolution since the European Renaissance – shook the foundations of physics in the first three decades of the twentieth century [1]. In the present era of global science, we often forget that this revolution was the handiwork of a small group of individuals who worked in a few places spread over a very limited region of Western Europe: London, Cambridge, Paris, Berlin, Göttingen, Munich, Bern, Copenhagen, Vienna . . . Most of these scientists knew each other well and often met at such venues as the Solvay conferences. Many of them also had the habit of exchanging their ideas through letters, which could typically reach the recipient in a day or two. Even in that era many decades before the internet and e-mails, the news of any major breakthrough would diffuse quickly through this closely-knit family of physicists. It certainly did appear that an outsider in a remote land had very little chance of making an inroad into this charmed circle. Apart from the fact that an outsider would be cut off from the intellectual atmosphere which nourished the minds of the insiders, the news of any important discovery would often reach a remote land only when the scientific journal reporting the discovery reached there by the slow surface mail many weeks after the publication of the work.

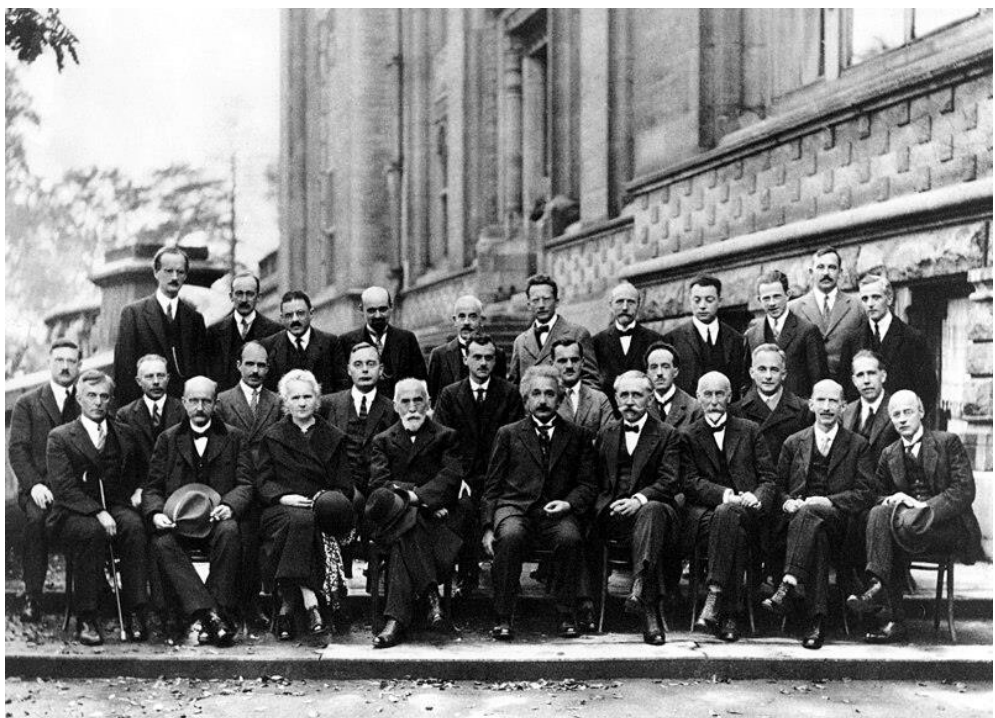

Figure 1.  Participants of 1927 Solvay conference.  Compton is seated in the middle row 4[th] from right (just behind Einstein).

When this scientific revolution was at its height, suddenly three important breakthroughs came within a few years from a distant impoverished land under the colonial rule without much of a previous tradition of research in modern physics: the Saha ionization equation in 1920 [2], the Bose statistics in 1924 [3] and the Raman effect in 1928 [4]. The three discoverers – Saha, Bose and Raman – were all faculty members of the small physics department at Calcutta University, newly established in 1916, although Bose had shifted to Dacca University by the time of his famous work.

Perhaps the Compton effect discovered in the USA in 1923 [5] was the only other discovery of that class pertaining to the new physics that was made outside Western Europe during this tumultuous decade of the 1920s. Even Compton was not so isolated. In the famous photo of the participants at the 1927 Solvay conference shown in Figure 1, you can see Compton seated just behind Einstein. Note that Saha and Bose had done their important works already a few years earlier, but they were not at this conference. The participants of the 1927 conference were almost like who's who in physics in that era. With the exception of Raman, Bose and Saha, virtually all physicists of their class at that time are in this photo. There is only one woman – Madame Curie – and no non-white person in the photo. Looking at this photo, one realizes how detached the Indian physicists were from the rest of the community of physicists who were creating the new physics.

Although Compton was not so isolated, still, working in another continent away from Western Europe, he probably appreciated the difficulties of Indian physicists quite acutely. He was always a champion of the Indian physicists. He nominated Saha for the Nobel Prize twice, though the nominations were not successful. When Raman was visiting the USA for the first time in 1924 – before he had discovered the Raman effect – Compton invited Raman for breakfast at his home. A rather awkward incident happened there. When Raman rang the doorbell, a maid opened the door and exclaimed: "Oh my God, it is a black man." Compton rushed there and profusely apologized to Raman [6].

During any epoch in history, we usually have a few scientific centres and some peripheral regions beyond these centres trying to catch up. It is often not realized how unique and unusual the physics contributions from India in the 1920s were in the annals of the history of science. It will be difficult to find another similar example from anywhere in the world in any branch of science: a succession of such extraordinary discoveries coming from a faraway peripheral land while a scientific revolution was at its peak. Such scientific creativity at the periphery away from scientific centres raises important broad questions in the history of science and the philosophy of science, which do not seem to have been analyzed much to this date [7-9].

There have been some studies of how the transfer of science takes place from the centres to the peripheries. Basalla in 1967 [10] gave an influential three-phase model of the spread of Western science outside the West. The era of Indian science we are discussing should presumably be identified with the middle phase which Basalla calls 'colonial science'. According to his model, this phase should have been preceded by a previous phase when European scholars explored the natural history of the country and should be followed by the last phase when the country becomes scientifically independent. This is a very generic model which glosses over the uniques aspects of the spread of Western science in different countries. Basalla's model and other similar models known to me seem rather inadequate in providing a framework for studying the sudden burst of physics creativity in colonial India.

How do we reconstruct the history of such unusual creative outbursts? Human creativity is something which always mystifies us. There are other examples of unusual simultaneous creativity. In classical Athens, three contemporaries Aeschylus, Sophocles and Euripides wrote tragedies which remained unsurpassed for nearly two millenia till Shakespeare wrote *Hamlet*. In the Renaissance Italy, three contemporaries Leonardo, Michelangelo and Raphael produced works of art which continue to amaze us. Nobody has been able to give a satisfactory explanation of why such creative geniuses sometimes appeared simultaneously in one place. The best one can do is to analyze the social conditions which might have been conducive for the sudden blossoming of such creativity. We shall now present a discussion of the social and intellectual climate of India from where these extraordinary physics discoveries came.

**The beginnings of Western science in India**

A large part of eastern India came under the colonial rule after the battle of Plassey in 1757. One responsibility of the colonial government was to look after the education of the 'natives' of the colony. For many centuries, India had some traditional centres of learning. There was a group of people – the 'Orientalists' – who favoured strengthening these traditional centres. On the other side were the 'Anglicists', who urged the government to introduce European style education. When the government decided to spend the education fund on Sanskrit schools, Rammohun Roy, often regarded as the father of modern India and a formidable Sanskrit scholar himself, wrote to the Governor-General in 1823:

> the Sanscrit system of education would be the best calculated to keep this country in darkness, if such had been the policy of the British legislature. But as the improvement of the native population is the object of the government, it will consequently promote a more liberal and enlightened system of instruction, embracing mathematics, natural philosophy, chemistry, anatomy, with other useful sciences . . .

The Anglicists finally became victorious when the famous *Minute on Education* written by Macaulay in 1835 was adopted as the government policy. India had a tradition of mathematics, astronomy and medicine from ancient times. It should be emphasized that the traditional Indian science did not expand to assimilate Western science, but was rather completely replaced by it! The scientists whom we are going to discuss were all trained exclusively in Western science. This was in contrast to what happened in literature and philosophy, where the Indian tradition and the European tradition continued to be taught side by side. The colonial government needed people who could control tropical diseases, explore the various natural resources which might be useful for the Empire and could construct railways, bridges and government buildings [11]. Scientific and technical education in India was completely geared towards fulfilling these practical needs of the Empire. There was no official support for fundamental research. It was often argued that a poor colony like India could not afford fundamental research.

To justify colonialism in Asia, the colonial masters had to invent an image of the 'Orient', as argued by Edward Said [12] and other post-colonial scholars. Orientals were supposed to be different from white people. Indians were thought to be spiritually inclined and not capable of practical things – ruling themselves or doing science. This idea provided the theoretical framework for the British to argue that they were in India for the benefit of the Indian people. Even many Indians in that era believed that Indians can never do scientific research – maybe due to their genetic constitution, or maybe the Indian diet was not suitable for the right kind of development of the brain. Even those who were more moderate held the widely accepted view that the Indian climate was not conducive for scientific research. This view about the influence of climate was so prevalent that I would like to give two quotations from two of the most famous professors teaching at Calcutta University in the 1920s. S. Radhakrishnan, who taught Indian philosophy, wrote in 1923 in support of the view that the physical conditions of India made Indians spiritually inclined [13]:

> Perhaps an enervating climate inclined the Indian to rest and retirement. The huge forests with their wide leafy avenues afforded great opportunities for the devout soul to wander peacefully through them, dream strange dreams and burst forth with joyous songs . . . It was in the . . . forest hermitage that the thinking men of India meditated on the deeper problems of existence.

On the other hand, C.V. Raman, when he was felicitated by Calcutta Corporation for being the first Asian to win a Nobel Prize in science, said [14]:

> Some have said that research work cannot be carried on successfully except in cool climates . . . A hot day in June is not an opportune moment to enter upon praise of the physical climate of Calcutta. But from the point of view of research, there is something more important than physical climate, and that is the intellectual climate of the environment. For a hundred years, . . . From Calcutta has gone forth a living stream of knowledge in many branches of study . . . It must be a profound privilege to work and live in such an environment.

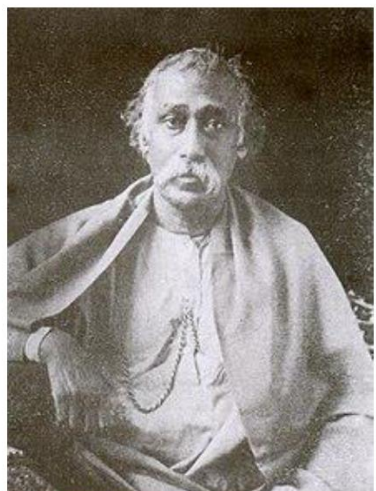 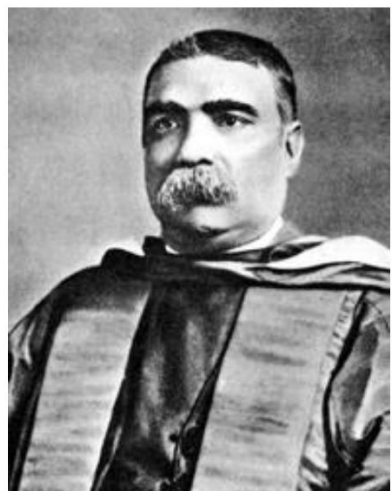

Figure 2. Mahendra Lal Sircar and Asutosh Mookerjee.

Indians who wanted India to undertake fundamental research in science knew that they could not get any government support. The first research institute of India – Indian Association for the Cultivation of Science (IASC) – was established in 1876 by Mahendra Lal Sircar (1833 – 1904). The initiative started when Sircar, a medical practitioner, wrote an article "On the desirability of a national institution for the cultivation of sciences by the natives of India" [15]. Sircar managed to obtain donations from Indians to construct a building and to set up a laboratory, but nobody came forward to do any research because there was nobody around trained to do scientific research. IACS would only arrange occasional lectures on scientific subjects. Sircar died a deeply disappointed man – thinking that all his efforts had gone in vain. As we shall discuss later, the first scientific work from Asia to win a Nobel Prize was done in Sircar's institute a few years after his death.

Jagadis Chandra Bose (1858 – 1937), who studied in Cambridge and then became physics professor at Presidency College Calcutta, was the first Indian to do original research in physics. Bose was the first man to produce microwaves in the laboratory in 1895, but then shifted to study plant physiology before the advent of the new physics. Unlike his great contemporary and colleague Prafulla Chandra Ray (1861 – 1944), who was the first Indian to do research in chemistry and who initiated a school of chemistry research in India by training many students, J.C. Bose did not train students to do research in physics. There was no continuity of a research tradition between him and the next generation of Indian physicists whom we are going to discuss. However, S.N. Bose and M.N. Saha were both students at Presidency College. J.C. Bose was their teacher, whom they respected deeply.

**Calcutta University and its Physics Department**

Let us now come to the history of Calcutta University, where the unusual physics discoveries were made. It was desirable to ensure some uniformity in the standards of different schools and colleges in the Empire. Three universities were established in the three Presidencies of the British empire – at Calcutta, Bombay and Madras – in 1857. We focus our attention on Calcutta University, which was initially a body for conducting examinations and granting degrees without having its own faculty until Asutosh Mookerjee (1864 – 1924) took over as Vice-Chancellor about half a century after its establishment. Mookerjee obtained fairly substantial donations from several wealthy Indians (lawyers like Taraknath Palit and Rashbehari Ghosh, as well as a few enlightened Maharajahs) to establish various academic departments and to start a few professorships. He laid down the foundation stone of the Science College of the University in 1914. Mookerjee himself was a brilliant mathematician, who had published a string of original papers in mathematics in leading journals at a very young age [16]. As there was very little scope of a career in mathematics in India of that time, Mookerjee had to take up the legal profession before being appointed to head Calcutta University. Although Mookerjee could not pursue mathematics research in his later life, he maintained a lifelong interest in mathematics and physics. He knew that it was an era of a revolution in physics and wanted to establish a Physics Department where there would be teaching and research in relativity and quantum theory. His well-wishers cautioned him that this was an absurd idea: there was nobody in the whole of India who even knew relativity and quantum theory.

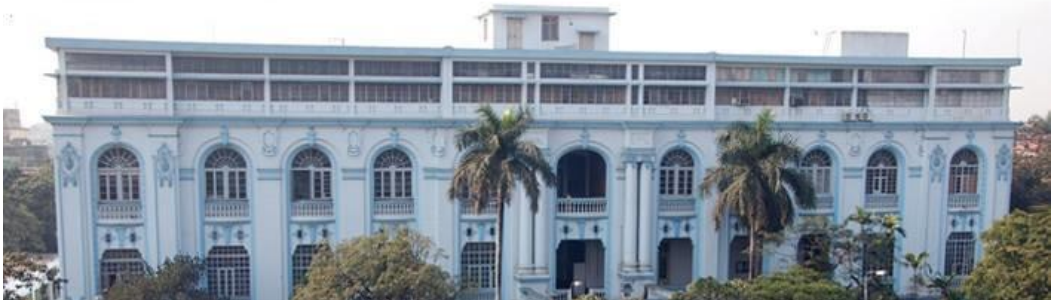

Figure 3. The Rajabazar Science College of Calcutta University, within which its historic Physics Department is still located.

The way Mookerjee built up the Physics Department can be the stuff of a fairy tale. He knew of a 26-year-old officer in the Finance Department, who was passionate about physics and had already published about a dozen papers in top international journals by carrying on research in his spare time. Mookerjee wanted to get him for the most prestigious chair of the fledgling Physics Department—the Palit Professorship. However, Mookerjee could offer him only Rs. 600 against his salary of Rs. 1100. Would he be willing to take up this professorship? The young man, C.V. Raman, jumped at the offer [17]. Mookerjee also needed younger persons to man the department. As it happened, the batch which completed master's degree in the year 1915 was an exceptional batch. Mookerjee called three bright boys of that batch for a discussion [18]. Only one of them, Sailen Ghose, was a student of physics. Although the other two, S.N. Bose and M.N. Saha, were students of mixed mathematics (what we would now call applied mathematics), Mookerjee knew that they were interested in physics. Mookerjee asked the three boys if they could teach the modern topics of physics which had never been taught in any Indian university so far. Saha was assigned to teach quantum theory and Bose was assigned to teach relativity. Sailen Ghose, who was a good experimenter, was given the job of designing the laboratory course and setting up the experiments. Before classes started in 1916, there was trouble. Although Sailen Ghose managed to acquire laboratory equipments and set up the laboratory, he could never formally join the department. He had connections with revolutionary groups fighting the British imperialism. Police found clues about this and raided his home when he was away. It appeared that he would be sent to the British penal colony of the Andaman Islands if caught. Ghose fled India in a ship bound for Philadelphia disguised as a Muslim crew member, thus putting down the curtains on what appeared to be a very promising career in physics.

Now we show a group photograph of the faculty members in the newly established Physics Department of Calcutta University taken around 1920. This happens to be the only existing photograph which has all three of our protagonists – S.N. Bose, C.V. Raman and M.N. Saha. Note that it was a rather small department consisting of only about a dozen faculty members.

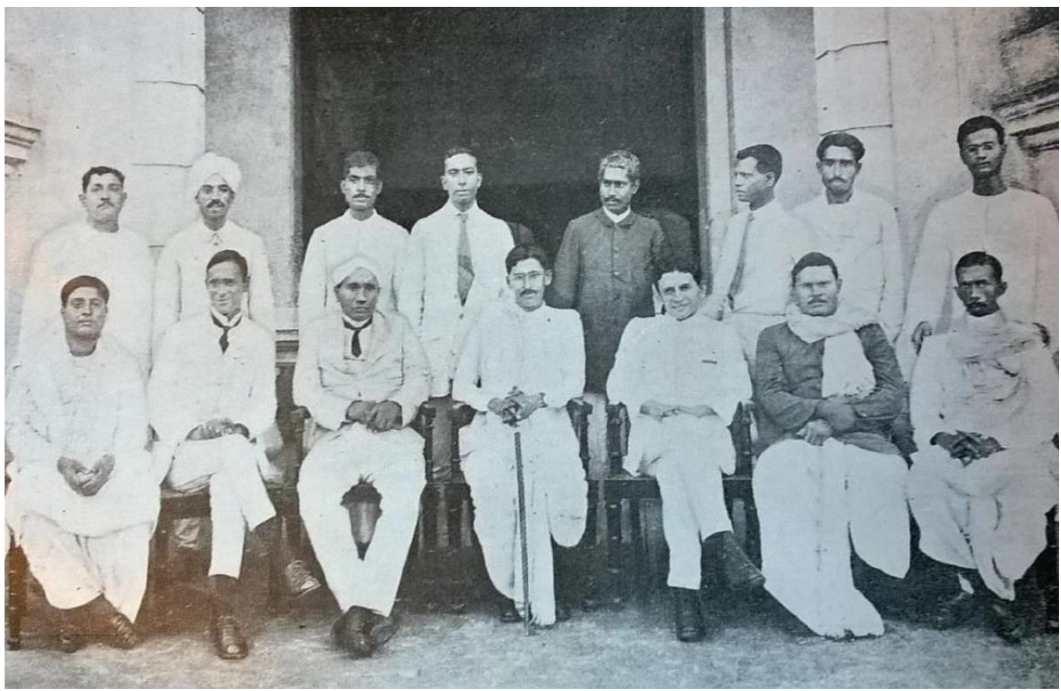

Figure 4. The physics faculty of Calcutta University around 1920. S.N. Bose, C.V. Raman and M.N. Saha are seated in the front row in the 1$^{st}$, 3$^{rd}$ and 4$^{th}$ positions from left.

Physics classes started in Science College in 1916 and, on 13 April 1919, a massacre took place in Jalianwala Bag, where several hundred unarmed men, women and children were gunned down. This was a turning point in the history of colonial India. Till that time, most of the Indian nationalists assumed that the British rule was to last in India for years to come. Their aim was to bring the shortcomings of the British rule to the attention of the colonial masters and to reform the system. Only after the Jalianwala Bag massacre, many started thinking for the first time that the colonial rule was an evil which must be overthrown. An extraordinary man came forward to lead perhaps the most extraordinary freedom movement in the history of the world. Mahatma Gandhi had returned to India from South Africa in 1915 and launched the non-violent non-cooperation movement. A new patriotic spirit pervaded the country in the 1920s: Indian scientists felt that they needed to show that they could also do what white men could do! C.V. Raman, who was seen weeping during the Nobel Prize ceremony, gave an eloquent account of it afterwards [19]:

> When the Nobel award was announced I saw it as a personal triumph. . . But when I sat in that crowded hall and I saw the sea of western faces surrounding me, and I, the only Indian, in my turban and closed coat, it dawned on me that I was really representing my people and my country . . . Then I turned around and saw the British Union Jack under which I had been sitting and it was then that I realized that my poor country, India, did not even have a flag of her own and it was this that triggered off my complete breakdown.

However, the freedom movement sealed the fate of Calcutta University, which was a government university depending on government funds even for paying many of the salaries. Many professors and students of Calcutta University were openly supportive of the freedom movement. The British government decided to punish the University by cutting funds, resulting in non-payment of professors' salaries for many months. Many of the famous professors, who could get jobs elsewhere, started leaving. Bose left for Dacca University in 1921 and Saha for Allahabad University in 1923. The mood of the era can be gauged from a letter dated 9 February 1921 which Vice-Chancellor Mookerjee wrote to Saha, who was on a sabbatical leave abroad when the financial crisis broke out [20]:

> I trust you will not hesitate to serve your Alma Mater when you return. I was deeply grieved to hear that Dr. Jnanendra Ghosh had decided to give up his Alma Mater and accept service in Dacca University. When will the children of our Alma Mater realise that it is absolutely necessary for all of them to stand by Her at the most critical period of the history of Her development?

You can almost hear the anguished voice of the great visionary, who built Asia's most extraordinary university of that era brick by brick and then saw it crumbling down. After Mookerjee suddenly died in Patna on 25 May 1924, there was no one to protect Calcutta University. Raman, the last of the famous physics professors, left in 1933 to become the first Indian Director of Indian Institute of Science in Bangalore. No fourth physics discovery of the same class has come out of India in the nearly one century which has elapsed since the golden era of the 1920s.

**Concluding Remarks**

Now I shall change my gear after this brief historical account and shall get into the discussion of physics in Part 2. Before coming to the discoveries of Saha, Bose and Raman, I shall have to give an account of the state of physics as a scientific discipline at that time. One question may come up at the very outset: Is it historiographically appropriate to discuss the science of Saha, Bose and Raman together? Scholars discuss science in ancient Greece, ancient China or ancient India, but usually not science in a modern nation in isolation. It is assumed that science in an ancient culture would develop without much interactions with the outside world and could be studied in isolation by itself. But that is not the case in the modern world. One honourable exception is Vol. 8 of *The Cambridge History of Science*, which has chapters on science of different nations in the recent past [21] – including a chapter on India by Deepak Kumar.

In the history of science, there is often a debate between internal history versus external history. Internal historians argue that the history of science should focus on the internal intellectual dynamics of science. On the other hand, external historians argue that the growth of science heavily depends on the surrounding social conditions and that should be the focus of historians. Certainly, scientists who faced identical obstacles to their scientific creativity can be treated together in external history. However, I argue that it is meaningful to treat the physics of Saha, Bose and Raman together even in an internal history. All of them worked on the general theme of the interaction between matter and radiation – the central theme in the physics of that era which led to the quantum revolution. In the beginning of Part 2, I shall say a few words about matter, then about radiation and finally about their interaction, before I discuss the discoveries of Saha, Bose and Raman.

**Acknowledgements.** I thank Sekhar Bandyopadhyay and Tulasi Parashar for urging me to give a seminar at the New Zealand India Research Institute at Victoria University of Wellington explaining the physics breakthroughs in colonial India to non-experts. Afterwards, I have given this seminar, on which this paper is based, in several other places. I thank Robert Anderson, Syamal Chakrabarti, Enakshi Chatterjee, Deepanwita Dasgupta, Subrata Dasgupta, Deepak Kumar, D.C.V. Mallik, Suprakash Roy and Rajinder Singh for many illuminating discussions on the history of science in colonial India over the years. I am grateful to K. Indulekha for going through the manuscript very carefully and for suggesting improvements.

**Bibliography**

[1] G. Gamow, *Thirty Years that Shook Physics* (Doubleday & Co, New York), 1966.
[2] M.N. Saha, Ionisation in the solar chromosphere, *Philosophical Magazize* **40,** 472-488, 1920.
[3] Bose, Plancks Gesetz und Lichtquantenhypothese, *Zeitschrift für Physik* **26**, 168-171, 1924.
[4] C.V. Raman and K.S. Krishnan, A new type of secondary radiation, *Nature* **121**, 501-502, 1928.
[5] A.H. Compton, A quantum theory of the scattering of X-rays by light elements, *Physical Review* **21**, 483-502, 1923.
[6] G. Venkataraman, *Journey into Light: Life and Science of C. V. Raman* (Indian Academy of Sciences, Bangalore), 1988, quote at p. 52.
[7] Somaditya Banerjee, *The Making of Modern Physics in Colonial India* (Routledge), 2020.
[8] Deepanwita Dasgupta, *Creativity from the Perhiphery* (University of Pittsburgh Press), 2021.
[9] G. Gangopadhyay, A. Kundu and R. Singh, *The Dazzling Dawn: Physics Department of Calcutta University (1916-36)* (Shaker Verlag, Düren), 2021.
[10] G. Basalla, The Spread of Western Science, *Science* **156**, 611-622, 1967.
[11] Deepak Kumar, *Science and Society in Modern India* (Cambridge University Press), 2023.
[12] Edward Said, *Orientalism: Western Concepts of the Orient* (Routlegde & Kegan Paul), 1978.
[13] S. Radhakrishnan, *Indian Philosophy, Vol. I* (George Allen & Unwin), 1923, quote at p. 22.
[14] S.N. Sen, *Prof. C.V. Raman: Scientific Work at Calcutta* (Indian Association for the Cultivation of Science), 1988, quote at pp. 270-271.
[15] M.L. Sircar, On the desirability of a national institution for the cultivation of sciences by the natives of India, *Calcutta Journal of Medicine* **2**, 286-291, 1869.
[16] Reena Bhaduri, *Sir Asutosh Mookerjee, Indian Association for the Cultivation of Science and Early Science Movement in India* (Asutosh Mookerjee Memorial Institute), 2013.
[17] Venkataraman, *Journey into Light*, p. 38.
[18] S.N. Bose, *Rachana Sankalan* (Bangiya Bigyan Parishad), 1980, pp. 226-227.
[19] S. Ramaseshan and C. Ramachandra Rao, *C. V. Raman: A Pictorial Biography* (Indian Academy of Sciences, Bangalore), 1988, quote at pp. 15-16.

[20] Dinesh Chandra Sinha, *Prasanga: Kolkata Biswabidyalay* (University of Calcutta), 2007, quote at p. 250.

[21] H.R. Slotten, R.L. Numbers and D.N. Livingstone (eds.), *The Cambridge History of Science. Volume 8: Modern Science in National, Transnational and Global Context* (Cambridge University Press), 2020.